\documentstyle[12pt,aps]{revtex}
\input epsf
\begin{document}
\title{
\hfill\parbox[t]{2in}{\rm\small\baselineskip 10pt
{~~~~~~JLAB-THY-98-23}\vfill~}
\vskip 0.01cm
\hfill\parbox[t]{2in}{\rm\small\baselineskip 10pt
{revised 28 August 1998}\vfill~}
\vskip 2cm
Valence Quark Spin Distribution Functions}

\vskip 1.0cm

\author{Nathan Isgur}
\address{Jefferson Lab\\
12000 Jefferson Avenue,
Newport News, Virginia 23606}
\maketitle

\vspace{4.0 cm}
\begin{center}  {\bf Abstract}  \end{center}
\vspace{.4 cm}
\begin{abstract}

 The hyperfine interactions of the constituent quark model
provide a natural explanation for many nucleon properties, 
including the $\Delta-N$ splitting,
the charge radius of the neutron, and the observation
that the proton's quark distribution function
ratio $d(x)/u(x) \rightarrow 0$ as $x \rightarrow 1$.
The hyperfine-perturbed quark model also makes predictions for
the nucleon spin-dependent distribution functions. 
Precision measurements of the resulting 
asymmetries $A_1^p(x)$ and $A_1^n(x)$ in the valence region can test this model and
thereby the hypothesis that the valence quark spin 
distributions are ``normal". 
\end{abstract}
\newpage
\section {Introduction}

	The quark model has enjoyed so much success as a qualitative guide to 
hadronic structure that the discovery that only about $30\%$ of the proton's spin
could be attributed to quark spin came as a surprise.  Since the
quark model remains unjustified within QCD, it is a misnomer to call
this ``proton spin surprise" the ``proton spin crisis".  However, whatever we call it, 
this result has generated much very productive experimental and theoretical
activity.

	While in general the spin of the proton could reside on any mixture of its quark
and gluon constituents or in their orbital angular momenta, 
a conservative interpretation \cite{GI} of the current situation is that the
valence quarks carry the spin expected by the quark model but that the low $x$ sea
of $q \bar q$ pairs is negatively polarized.  In this case 
$\Sigma$ (defined to be
twice the expectation value of the 
quark plus antiquark spin along the spin direction of
a polarized proton, so that $\Sigma = 1$ would saturate the proton spin),
when decomposed into its valence and sea components, would be
\begin{equation}
\Sigma = \Sigma_{v} + \sum_q \Delta (q+\bar q)_{sea}
\end {equation}
where $\Sigma_{v} = \int dx \Sigma_{v} (x)$ is twice the spin on the valence quarks and
$\Delta q_{sea} = \int dx \Delta q_{sea}(x)$
and $\Delta \bar q_{sea} = \int dx \Delta \bar q_{sea}(x)$ are, respectively, 
twice the spin on the sea quarks and 
antiquarks of flavor $q$.  If the valence quarks were in nonrelativistic
$S$-waves as in the naive quark model, then 
$\Sigma_{v}$ would be unity.  However, as has been appreciated for nearly thirty
years \cite{gA}, in realistic valence quark models lower components of quarks spinors
convert about $25 \%$ of the quark spin into orbital angular momentum so that
$\Sigma_{v} \simeq 0.75$.  If in addition each of the three light quark flavors 
carries $\Delta (q+\bar q)_{sea} \simeq -0.15$, a very modest per flavor 
effect, $\Sigma \simeq 0.30$ would follow.
Sea quark polarizations of just this sign and magnitude have recently been obtained in a 
realistic model of $q \bar q$ pair creation \cite{GI}.  (In a more general context,
such small $\Delta (q+\bar q)_{sea}$ values are perfectly consistent with a ${1 / N_c}$ 
expansion of QCD (where
sea quarks appear at order ${1 / N_c}$ {\it via} quark-antiquark loops).  
Note that the condition  $\Delta (q+\bar q)_{sea} \ll 1$,
not how accurately $\Sigma_{v}$
approximates unity, determines the applicability
of the $1 / N_c$ expansion: any nonzero $\Delta (q+\bar q)_{sea}$ would lead to a ``spin crisis" as $N_{f}$
(the number of light flavors) tends to infinity. )

	In the conservative scenario just described, both the $ 25 \%$ relativistic 
quenching of spin from $\Sigma _{v}$ and the negative polarization of $\Delta (q+\bar q)_{sea}$ 
are compensated by orbital angular momentum.  In general, 
however, we are only guaranteed that
\begin {equation}
\Sigma_{v} + \sum_q \Delta (q+\bar q)_{sea} + 2L_{q} + \Sigma_{g} = 1
\end {equation}
(where $L_{q}$ is  the quark and antiquark 
orbital angular momentum and ${1 \over 2} \Sigma _{g}$ is the total
angular momentum residing in the gluonic fields), so major experimental efforts are planned to 
measure the component parts of Eq. (2) in an effort to disentangle the ``spin crisis".  
These efforts begin with planned extensions of deep inelastic lepton scattering
measurements of the proton and neutron spin
structure functions down to very small $x$ to complete the integrals required to calculate
$\Sigma$, and studies of the $Q^2$-dependence of
spin structure functions to make inferences about $\Delta g (x)$, the gluon helicity
contribution to $\Sigma_{g} (x)$.  Major
efforts are also planned to directly measure $\Delta g (x)$ based on helicity-dependent 
gluon-parton cross sections.  In addition to these classical inclusive measurements, 
flavor-tagging semi-inclusive experiments 
are planned to measure separately 
$\Delta s_{sea}{(x)}$, $\Delta \bar s_{sea}{(x)}$,  
$\Delta \bar u_{sea}(x)$,
$\Delta \bar d_{sea}(x)$, and also 
the quark contributions 
$\Delta u (x) \equiv \Delta u_{v}(x)+ \Delta u_{sea}(x)$ and 
$\Delta d (x) \equiv \Delta d_{v}(x)+ \Delta d_{sea}(x)$.  
(Note that it is not  possible 
to experimentally separate the quark contributions $\Delta u_{sea}(x)$ and $\Delta d_{sea}(x)$ from 
$\Delta u_v(x)$ and $\Delta d_v(x)$:  this separation is  conceptual only.)  
Additional complementary information on the $s \bar s$ content of the proton is expected from
planned measurements of the electric and magnetic form factors $G^s_E$ and
$G^s_M$ of the $\bar s \gamma^{\mu} s$ current using parity-violating electron-nucleon elastic scattering.

	{\it Given the substantial effort being devoted to this problem, it is surprising 
that we still do not know whether our original simple picture of the spin structure of the 
valence quarks is right!}	To some degree this is because this question is not well-defined:
in contrast to other methods ({\it e.g.}, QCD sum rules \cite{Ioffe}), the quark
model is not normally embedded in a field-theoretic framework. As a result,
there are many difficulties in making comparisons between the ``predictions" of the quark
model and the precisely defined quantities measured in deep inelastic scattering. As 
two illustrations of such difficulties, I note that:
1) the separation of Eq. (2) is $Q^2$-dependent ({\it e.g.}, $\Delta g$ might be small at 
low $Q^2$ but very important at large $Q^2$) and, as mentioned above,
2) the  $u$ and $d$ contributions to $\Delta u_{v} (x)$ and
$\Delta d_{v} (x)$ cannot be disentangled from those to $\Delta u_{sea} (x)$ and $\Delta d_{sea} (x)$.
 However, beyond $x \simeq 0.3$, 
sea quarks and antiquarks are scarce and, since gluons are too,
such intrinsically field-theoretic issues as the factorization scheme dependence of
$\Sigma_v(x)$ associated with the gluon anomaly \cite{anomaly} may be neglected.  
Thus while the integral values 
$\Delta u_{v} = \int dx \Delta u_{v} (x)$ and $\Delta d_{v} = \int dx \Delta d_{v} (x)$ cannot be
checked, those fractions of the distributions $\Delta u (x)$ and $\Delta d (x)$ extending
beyond $x \simeq 0.3$ may be compared with valence quark model expectations
with a residual ambiguity associated only with their $Q^2$ evolution. Although in what follows I will
imagine distribution functions devolved
to the ``quark model scale" $Q_{0} ^{2} \simeq 1$ GeV$^{2}$, given
this residual ambiguity I will avoid predictions of the $x$-dependence of
distribution functions and focus instead on the polarization
asymmetries $A_{1}^{p} (x)$ and $A_{1}^{n} (x)$ which depend
only on ratios of distribution functions and which should therefore have minimal $Q^2$-dependence.
It is unfortunate that the
current experimental situation for $A_{1}^{p} (x)$ and $A_{1}^{n} (x)$ 
for $x>0.3$ leaves much to be desired (see Figs. 1): 
it is even consistent with the
naive $SU(6)$ predictions.
	
     What are the valence quark model predictions for 
the resulting polarization asymmetries $A_{1}^{p} (x)$ and $A_{1}^{n} (x)$ in the
valence region?  Ignoring $Q ^{2}$ evolution, they are
\begin {equation}
A_{1}^{p}(x) = {{4 \Delta u_{v} (x) + \Delta d_{v} (x)} \over {4 u_{v} (x) +
d_{v} (x)}}
\end {equation}
\begin {equation}
A_{1} ^{n} (x) = {{4 \Delta d_{v} (x) + \Delta u_{v} (x)} \over {4 d_{v} (x) + u_{v} (x) }}
\end {equation}
\noindent
where $u_{v} (x)$ and $d_{v} (x)$ are the unpolarized valence distribution functions which integrate
to $2$ and $1$, respectively.  From these    
formulas it is clear that the predictions depend on knowing the interplay between the
valence quark spin and momentum wavefunctions so that there can be no 
unique prediction of the valence quark model for these asymmetries. However, 
I will argue here that its predictions are sufficiently well-determined
that they can be used to answer the simple question of whether the valence
spin structure is ``normal" or not. 

   Aside from this observation, there is little in this paper that could not be
extracted from earlier work on this subject to which I will refer below. However, 
the results of this earlier work vary widely since they are based on diverse methods
of dealing with relativistic internal quark motion, various prescriptions
for boosting to the infinite momentum frame, {\it ad hoc} versus dynamical
origins for the assumed $SU(6)$-breaking, potential {\it versus} bag models, and
choices of quark masses. Here I will {\it assume} that the hyperfine interaction 
is responsible for the $d(x)/u(x)$
ratio as $x \rightarrow 1$, and then normalize predictions 
for the valence quark spin distribution functions to the data
on this ratio. In doing so, I will not only avoid much model dependence, but also 
most of the pitfalls discussed above associated with not knowing precisely how 
to embed the quark model in field theory.
\break

%
%
\begin{center}
~
\epsfxsize=3.0in  \epsfbox{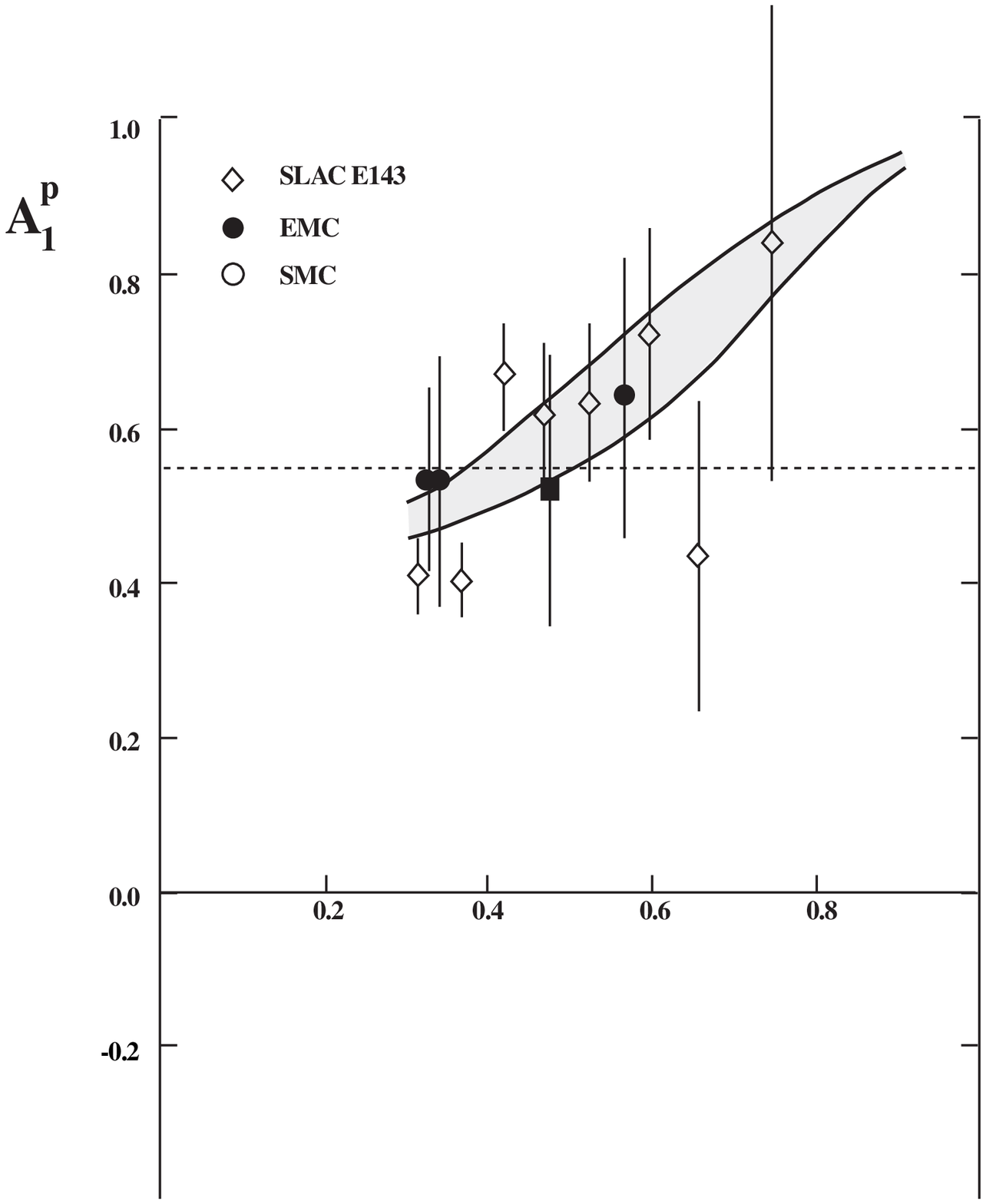}
\vspace*{0.1in}
~
\end{center}

\noindent{ Figure 1(a): Data \cite{data} on $A_{1}^{p}$ and the prediction (shaded band)
of the model described in the text; the $SU(6)$ prediction is ${5 \over 9}$ (dotted line).}

\bigskip
%
%
\begin{center}
~
\epsfxsize=3.0in \epsfbox{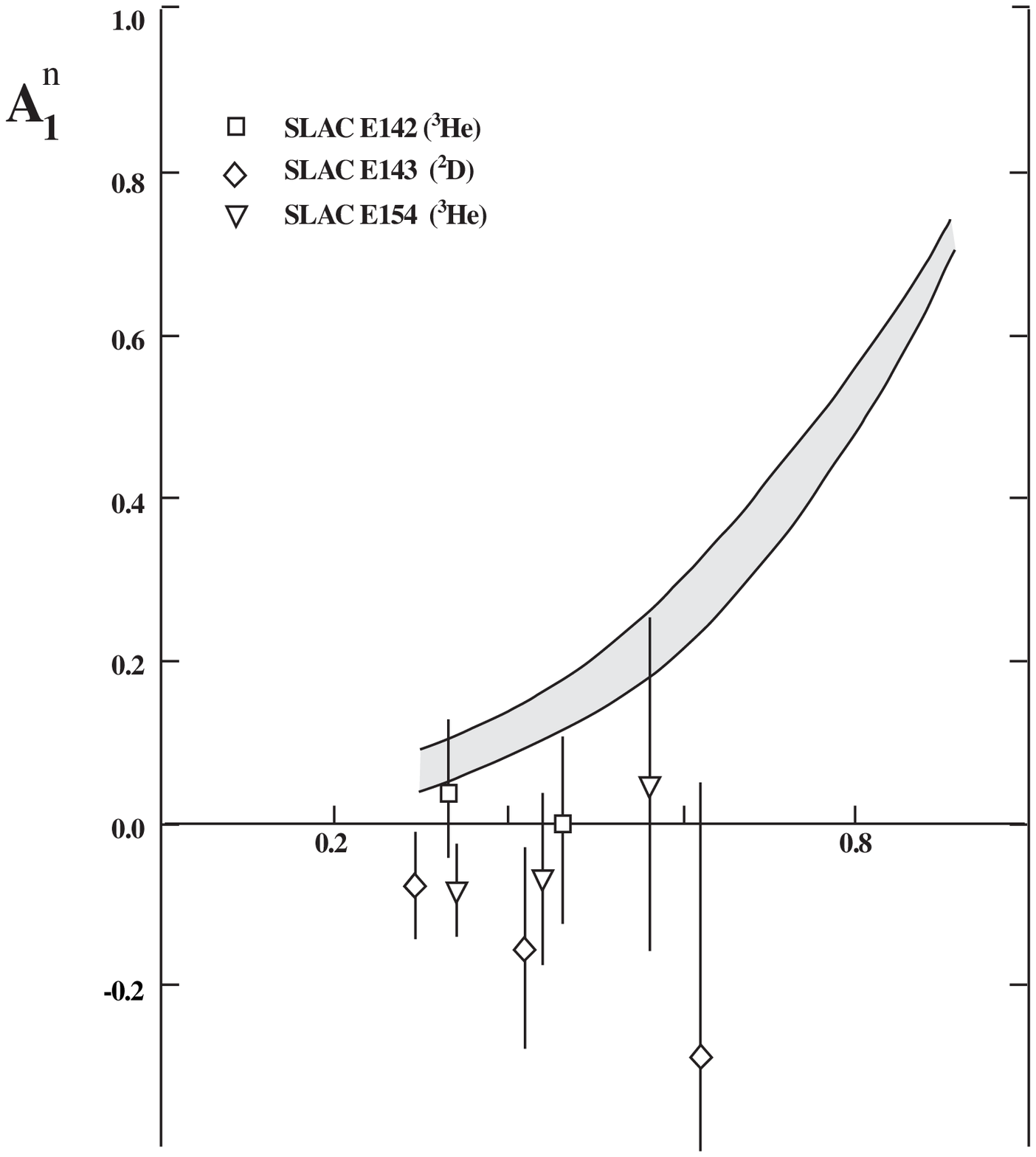}
\vspace*{0.1in}
~
\end{center}

\noindent {Figure 1(b): Data \cite{data} on $A_{1}^{n}$  and the prediction (shaded band)
of the model described in the text; the $SU(6)$ prediction is $0$.}

\bigskip

    The body of this paper builds up to these predictions
in steps. In the next Section I will review the naive $SU(6)$ predictions and
then modify them within the context of
$SU(6)$ by allowing the quarks to have relativistic internal
motions. I then describe the breaking of $SU(6)$-symmetric quark spin distributions
in the hyperfine-perturbed quark
model and close with a
brief historical overview. 

\section {The $SU(6)$ and ``Relativized  $SU(6)$" Distribution Functions}

	I begin by recalling that in  $SU (6)$ one may simply write
\begin{equation}
p \uparrow = uud C ^{A} \psi ^{S} \chi _{+} ^{\lambda}
\end{equation}
\begin{equation}
n \uparrow = ddu C^{A} \psi ^{S}  \chi_{+}^{\lambda}
\end{equation}
where since $C^{A}$ and $\psi^{S}$ are the antisymmetric color and symmetric $L=0$ spatial wavefunctions, 
\begin{equation}
\chi_{+}^{\lambda} = -\sqrt{1 \over 6}  (\uparrow\downarrow\uparrow +
\downarrow\uparrow\uparrow - 2\uparrow\uparrow\downarrow)
\end{equation}
is the unique spin-${1\over 2}$
wavefunction which is symmetric in the first two quarks as required by the Pauli principle 
\cite{uds}.  In the nonrelativistic $SU(6)$ quark model one therefore expects
\begin {equation}
u_{v} \uparrow (x) = {5\over 3} v_{SU (6)} (x)
\end {equation}
\begin {equation}
u_{v} \downarrow (x) = {1\over 3} v_{SU (6)} (x)
\end {equation}
\begin {equation}
d_{v} \uparrow (x) = {1\over 3} v_{SU (6)} (x)
\end {equation}
\begin {equation}
d_{v} \downarrow (x)= {2\over 3} v_{SU (6)} (x)
\end {equation} 
\medskip
where $v_{SU(6)}(x)$ is the universal $SU(6)$ distribution function associated
with $\psi^S$. These distributions 
lead to the standard $SU(6)$ predictions $d(x)/u(x)=1/2$,
$A_1^p(x)=5/9$ and $A_1^n(x)=0$, 
and $G_A=5/3$.
When the relativistic quenching mentioned above \cite{gA}
is turned on, it creates an $x$-dependent probability which
we denote by ${1 \over 2} c_A(x)$ for a spin up (down) quark 
to be flipped to down (up). This reshuffling of probability
leads to the ``relativistic $SU(6)$" spin distributions
\begin {equation}
u_{v} \uparrow (x) = [{5\over 3}-{2\over 3}c_A(x)] v_{SU (6)} (x)
\end {equation}
\begin {equation}
u_{v} \downarrow (x) = [{1\over 3}+{2\over 3}c_A(x)] v_{SU (6)} (x)
\end {equation}
\begin {equation}
d_{v} \uparrow (x) = [{1\over 3}+{1\over 6}c_A(x)] v_{SU (6)} (x)
\end {equation}
\begin {equation}
d_{v} \downarrow (x)= [{2\over 3}-{1\over 6}c_A(x)] v_{SU (6)} (x)
\end {equation} 
where with the nucleon expectation value $\langle 1-c_{A}(x) \rangle_N \simeq {3\over 5} G_{A} \simeq 0.75$,
the integrated valence spins become 
\begin {equation}
\Delta u_v \simeq +{4 \over 5} G_{A}
\end {equation}
\begin {equation}
\Delta d_v \simeq -{1 \over 5} G_{A}~,
\end {equation}
\medskip
so that the ``relativistic $SU(6)$" spin distributions 
satisfy the Bjorken sum rule. However, among other problems, the
``relativistic $SU(6)$" model 
still makes the incorrect prediction
$d(x)/u(x)=1/2$. Note that
the model also predicts that
$A_1^p(x)=[1-c_A(x)]5/9$ and $A_1^n(x)=0$ as $x \rightarrow 1$, which, 
since $c_A(x) \rightarrow 0$ as $x \rightarrow 1$, 
is not obviously wrong (see Fig. 1).

\section {Predictions of the Hyperfine-Perturbed Quark Model}

Since the zeroth-order nucleons are pure $S$-waves, in 
the hyperfine-perturbed quark model  \cite{IKquarkmodel},  only the Fermi
contact part of the hyperfine interaction (the 
$\vec S_i \cdot \vec S_j \delta^3(\vec r_{ij})$ force
responsible for the $\Delta - N$ mass splitting)
is operative in perturbing the nucleon's energy in first order.
What does this perturbation do?  In the nucleon rest frame,
quark pairs with spin 1 have their energies raised (as in the $\Delta$) while pairs with spin zero
have their energies lowered.  Since $\chi^{\lambda}$ has the two $u$ quarks in a pure spin one state,
while each $ud$ pair is in a mixture of spin one and spin zero (with spin zero dominant so that the 
net perturbation in a nucleon decreases its energy), up quarks acquire higher average energy than down
quarks.  This physics then immediately suggests that the neutron will have a negative charge radius
and that $d(x)/u(x)$ will vanish as $x \rightarrow 1$
\cite{Carlitz,IKK,zakopane} . Since the individual spin components 
of $\chi^{\lambda}$ are {\it not} in an eigenstate
of the hyperfine interaction, it is less obvious what the effects are on the spin-dependent
distribution functions. 

   These effects are encoded in the $L=0$ component of the hyperfine-perturbed
wavefunction 
\begin {equation}
u u d C^{A} \Biggl[cos {\theta}_m \psi^{S}\chi^{\lambda}_+ + sin {\theta}_m
\sqrt{1\over 2}(\psi^{\rho} \chi^{\rho}_{+}-\psi^{\lambda} \chi^{\lambda}_+)\Biggr]
\end {equation}
\medskip
where $(\psi^{\rho},\psi^{\lambda})$ are mixed symmetry wavefunctions of the permutation 
group $S_{3}$ which are 
antisymmetric ($\rho$) and symmetric ($\lambda$) under $1 \leftrightarrow 2$ interchange, where
\begin {equation}
\chi_{+}^{\rho} = \sqrt{1 \over 2} (\uparrow \downarrow - \downarrow \uparrow) \uparrow ~~~,
\end {equation}
\medskip
and where $\theta_m$ is a  small mixing angle induced by $SU (6)$-breaking
interactions \cite{IKK}.  (Since, as explained above, the $L = 0$ 
ground state energies are perturbed in first order only by the 
$\vec S_i \cdot \vec S_j$ interaction, one can ignore $L=2$ and totally antisymmetric $L=0$ admixtures.)  It follows that the rest frame 
probabilities of spin up and spin down
$d$ quarks are, to first order in $\theta_m$,
\begin {equation}
P(d\uparrow) =  {1 \over 3}\biggl| \psi^{S} - 
\sqrt{1\over 2} \theta_m\psi^{\lambda}\biggr|^{2}
\end {equation}
\begin {equation}
P(d \downarrow) = {2 \over 3}\biggl| \psi^{S} - 
\sqrt{1\over 2} \theta_m\psi^{\lambda}\biggr|^{2},
\end {equation}
since the $\psi^{\rho} \chi_{+}^{\rho}$ piece of the wavefunction 
does not interfere with the other terms in the probability distribution.
In these formulas I have suppressed coordinate labels
which indicate that the probability $P(d \uparrow)$
($P(d \downarrow)$) is that for finding a spin up (spin down) $d$ 
quark at a point $\vec r_d$ while the
two up quarks are at positions $\vec a$ and $\vec b$.

Similarly one finds
\begin {equation}
P(u \uparrow) =  {5 \over 3}\biggl| \psi^{S} - 
\sqrt{1\over 2} \theta_m\psi^{\lambda}\biggr|^{2}
-\sqrt{2\over 3}\theta_m\psi^S\psi^{\rho}
\end {equation}
\begin {equation}
P(u \downarrow) = {1 \over 3}\biggl| \psi^{S} - 
\sqrt{1\over 2} \theta_m\psi^{\lambda}\biggr|^{2}
+\sqrt{2\over 3}\theta_m\psi^S\psi^{\rho}
\end {equation}
where now the wavefunction $\psi^{\rho}$ does play a role. 
I have now suppressed coordinate labels
which indicate that the probability $P(u \uparrow)$ 
($P(u \downarrow)$) is that for finding a spin up (spin down) $u$ 
quark at a point $\vec r_u$ while the
other up quark is at position $\vec \alpha$ and the $d$ quark is
at position $\vec \beta$.

    Note that, as advertized, the net leading-order effect of the $SU(6)$-breaking 
in the spin-averaged probabilities is to create distributions of mixed
symmetry that allow the $d$ quark to have a different
probability distribution
from the two $u$ quarks.   
With the calculated quark model value \cite{IKK} $sin \theta_m \simeq -0.23$, the 
distortion of the $SU(6)$-symmetric probabilities is substantial.
I now make the natural assumption that this distortion
translates into the observation that  $d(x)/u(x) \rightarrow 0$ as $x \rightarrow 1$, 
and associate the measured $u(x)$ and $d(x)$ with functions $u_v(x)$ and $d_v(x)$ 
associated with the spin-averaged probability $\biggl| \psi^{S} - 
\sqrt{1\over 2} \theta_m\psi^{\lambda}\biggr|^{2}$.
This remarkably simple picture then leads to the ``standard" prediction
$F_{2}^{n}/F_{2}^{p} \rightarrow {1\over 4}$ as $x \rightarrow 1$.

    The predictions of Eqs. (20)-(23) for $A_1^p(x)$ and $A_1^n(x)$ may 
easily be deduced using the properties of the
mixed symmetry pair of wavefunctions ($\psi^{\rho}$, $\psi^{\lambda}$)
under the permutation group $S_3$. Such an analysis reveals
that the hyperfine interactions have distorted the distributions
of $u\downarrow$, $d\uparrow$, and $d\downarrow$ identically, and that
the entire dominance of $u$ quarks as $x \rightarrow 1$ is due to 
$u\uparrow(x)$. This means that
\begin {equation}
u_{v} \uparrow (x) = [1-{1\over 2}c_A(x)] u_v (x) -  {1\over 3}[1-c_A(x)]d_v (x)
\end {equation}
\begin {equation}
u_{v} \downarrow (x) ={1\over 3}[1-c_A(x)]d_v (x)+ {1\over 2}c_A(x) u_v (x)
\end {equation}
\begin {equation}
d_{v} \uparrow (x) = {1\over 3}[1+{1\over 2}c_A(x)] d_v (x)
\end {equation}
\begin {equation}
d_{v} \downarrow (x)= {2\over 3}[1-{1\over 4}c_A(x)] d_v (x)
\end {equation} 
The resulting predictions for $A_1^p(x)$ and $A_1^n(x)$
in the valence region, shown in Fig. 1,
can be obtained without engaging in
an elaborate parameterization of structure functions.
Using the rough parameterizations $d(x)/u(x) \simeq {\kappa}(1-x)$
as $x \rightarrow 1$ (with $0.5< \kappa < 0.6$) and 
$c_A(x)=nx(1-x)^n$ (which builds in $c_A(x) \rightarrow 0$ as $x \rightarrow 1$
and $x \rightarrow 0$ and for
$2<n<4$ gives the required quenching of $G_A$), produces the narrow bands shown in the figure.
As $x \rightarrow 1$ both $A_{1}^{p}$
and $A_{1}^{n}$ tend to $1$, but I show the predictions only in the region where 
the valence quark wavefunction is large since
very small effects might become important at the endpoint \cite{Farrar}.

\section{Some History}
\bigskip
    The history of the prediction of the effects of $SU(6)$-breaking
on the quark distribution functions 
in the valence region is somewhat convoluted. It perhaps begins with the
parton model discussion by 
Feynman \cite{Feynman}  who argues 
that as a $u$ or $d$ quark approaches $x=1$, it must leave behind ``wee" 
partons with either $I=0$ or $I=1$,
and that these two configurations are unlikely to have the same $x$-dependence. 
He then notes that if the $I=0$
configuration dominates as $x \rightarrow 1$, the observed ratio $F_2^n/F_2^p=1/4$ would follow. 
If we take the modern view that this high $x$ behaviour will be controlled
by the valence quarks, and note the quark model
correlation between isospin and spin in the valence quark
sector, this argument would also naively lead to the conclusion
that $u _{v} \uparrow (x)$ will dominate as $x \rightarrow 1$.
While correct, since Feynman's
argument relies on the ``wee" partons being uncorrelated with the leading quark, and 
so does not take into account the required antisymmetrization
between the leading $u$ quark and the ``wee" $u$ quark, its predictions for the full
valence region are unclear.

   A more complete quark model argument is given in the papers 
of Close \cite{Close} and Carlitz and Kaur \cite{Kaur}.
They argued that $SU(6)$-breaking
changes Eq. (5) into 
\begin {equation}
p \uparrow = uud C_{A} [{\sqrt 3 \over 2} \uparrow \chi_{ud}^{0}\psi^0 + {1 \over 2} (\sqrt {1 \over 3}
\uparrow \chi_{ud}^{10} - \sqrt{2 \over 3} \downarrow \chi_{ud}^{11})\psi^1]
\end {equation}
\medskip
where $\chi_{ud}^{0} = \sqrt {1 \over 2} (\uparrow\downarrow -\downarrow\uparrow)$, 
$\chi_{ud}^{11}= \uparrow \uparrow$, and $\chi_{ud}^{10} = \sqrt{1 \over 2} 
(\uparrow \downarrow + \downarrow \uparrow)$ are the $S=0$ and two $S=1$ $ud$ spin wavefunctions.  For
$\psi^{0} = \psi^{1} = \psi^{S}$, this wavefunction collapses to (5). Referring
to hyperfine forces as driving the physics (which is equivalent
to Feynman's assumption in this case), these papers
posit that
$SU(6)$-breaking leads to $\psi^{0} \not= \psi^1$, which would in turn lead to the
relations
$u_{v} \uparrow (x) = {1\over 6}v_1(x)+{3\over 2}v_0(x)$,
$u_{v} \downarrow (x) = {1\over 3}v_1(x)$,
$d_{v} \uparrow (x) = {1\over 3}v_1(x)$,
and $d_{v} \downarrow (x)= {2\over 3}v_1(x)$ 
in terms of distribution functions $v_0$ and $v_1$ associated with $\psi^0$
and $\psi^1$, respectively,
with $v_1/v_0 \rightarrow 0$ as $x \rightarrow 1$.  This model thus also leads to 
$F_{2}^{n}/F_{2}^{p} \rightarrow {1\over 4}$ as $x \rightarrow 1$ and it
predicts that both $A_1^p(x)$ and $A_1^n(x) \rightarrow 1$ as $x \rightarrow 1$.
While
assuming $\psi^{0} \not= \psi^1$ is very natural, since these diquark spin states
are eigenstates of the hyperfine interaction, this assumption is
not consistent with the Pauli principle unless $\psi^0 $ and $\psi^1$ have
very special properties under the permutation
group $S_3$ or the wavefunction (28) is antisymmetrized. 
Thus, as with Feynman's argument, it is unclear what these models predict
for the full valence region. The
closely related model
of Close and Thomas \cite{CloseThomas} is based on examining the energy of the spectator diquark {\it after}
the deep inelastic scattering in a rest frame calculation
of deep inelastic structure functions.  Although, as pointed 
out by the authors, their calculation suffers from the fact
that the diquark is a colored object which cannot have a well-defined energy,
this calculation emphasizes the same physics and reaches the same conclusions
as Refs. \cite{Close,Kaur}. Given that the impact of the hyperfine interaction
is implemented somewhat intuitively in this work, it is once again unclear
whether the results presented are reliable for anything other than the
$x \rightarrow 1$ behaviour.

   Although they do not use the hyperfine-perturbed
quark model, the formalism required to
deal explicitly with the fully antisymmetrized
nucleon wavefunction seems to have first been applied to the valence quark
spin distribution functions by Le Yaouanc {\it et al.} \cite{LeY}.  They introduce an $SU(6)$
intraband mixing between the ground state $[56,0^+]$ and
the mixed symmetry $[70,0^+]$ in an attempt to account for the observed 
behaviour $F_{2}^{n}/F_{2}^{p} \rightarrow {1\over 4}$ as $x \rightarrow 1$, {\it i.e.},
$d(x)/u(x) \rightarrow 0$ as $x \rightarrow 1$.  This is precisely the 
kind of mixing introduced in (18) as required by color hyperfine interactions. 
(In fact, using this formalism makes calculations much simpler
than in the $uds$ basis, though perhaps less physically transparent.) They then
make a prescription to boost this mixed wavefunction into the infinite momentum frame, 
fit the mixing angle to the data, and discuss the implications of such mixing to a wide range of phenomena.

   More recently, a number of authors \cite{Dziembowski,McCarthy,Goldman,Weber}
have addressed the connection between
the hyperfine-perturbed quark model (either potential-based or bag-like) and 
the quark distribution functions. Most of these papers find the same
two key effects I have emphasized here: axial current quenching
by internal quark motion and u quark dominance as $x \rightarrow 1$.

   Despite this extensive body of work \cite{Feynman,Close,Kaur,CloseThomas,LeY,Dziembowski,McCarthy,Goldman,Weber},
it does not seem to be widely appreciated that the hyperfine-perturbed valence quark model
makes quite clear predictions for the asymmetries $A_1^p(x)$ and $A_1^n(x)$ in the valence region. I attribute
this state of affairs to the fact that this work  has been very ambitious:
most authors have attempted ``absolute" calculations
of structure functions. In doing so they encountered many
obstacles, which forced them to a variety of assumptions,
approximations, and ``procedures". The result is a wide range
of predictions for the structure functions with apparent agreement {\it only} on their
qualitative features.

\section{CONCLUSIONS}
\bigskip

   In this paper I have shown that once it is assumed that
the hyperfine perturbations of the quark model are responsible for the $SU(6)$-breaking
observed in the structure functions, a very narrow band of predictions follows.
In a broader context, I have argued
that the extensive measurements and theoretical studies engendered
by the ``spin crisis" should be anchored in knowledge of whether the valence quark spin
distributions are in fact anomalous.   Thus whether the distributions 
described here prove to
be correct when confronted with the data will be interesting, but
not as important as the fact that such data
will indicate whether the valence spin structure functions
are in fact anomalous, and thus guide the search for where 
the resolution of the ``spin crisis" is to be found.

\begin{acknowledgements}

    I would like to thank Frank Close, Claudio Coriano, Lionel Gordon, Gordon Ramsey,
and Tony Thomas for patiently discussing with me the theory behind an early incorrect draft of this paper,
and Gordon Cates, Bernard Frois, Krishna Kumar, Jim McCarthy, and Paul Souder
for encouraging me to define the quark model predictions for the 
valence spin structure functions. I am particularly grateful to Zein-Eddine Meziani
who rekindled my interest in this problem by explaining to me that precision measurements
of these structure functions are possible with
CEBAF at Jefferson Lab.

\end{acknowledgements}

\end{document}